\documentclass[cup5b]{caps}

\def\etal{{et\ al.\ }}
\def\aa{{A\&A}}

\def\aj{{AJ}}
\def\annrev{{ARA\&A}}
\def\apj{{ApJ}}
\def\apjl{{ApJL}}
\def\apjs{{ApJS}}
\def\baas{{BAAS}}
\def\mnras{{MNRAS}}
\def\nat{{Nature}}
\def\pasp{{PASP}}
\def\df{DF}

\def\kms{{\rm km\, s}^{-1}}
\newcommand{\msun}{M_{\odot}\,}
\def\ltorder{\mathrel{\hbox{\rlap{\hbox{%
 \lower4pt\hbox{$\sim$}}}\hbox{$<$}}}}
\def\gtorder{\mathrel{\hbox{\rlap{\hbox{%
 \lower4pt\hbox{$\sim$}}}\hbox{$>$}}}}

\usepackage{graphicx}
\usepackage{amssymb}
\usepackage{ociwsymp1}
\HeadText{S. Sigurdsson}

\def\plotone#1{\centering \leavevmode
\includegraphics[width=.95\columnwidth]{#1}}

\def\plottwo#1#2{\centering \leavevmode
\includegraphics[width=.45\columnwidth]{#1} \hfil
\includegraphics[width=.45\columnwidth]{#2}}

\begin{document}

\pagenumbering{arabic}

\author[]{S. SIGURDSSON\\
%525 Davey Laboratory,\\ Department of Astronomy \& Astrophysics, \\
%Pennsylvania State University,\\ University Park, Pa 16802, USA
Department of Astronomy \& Astrophysics, \\
Pennsylvania State University
}

\chapter{Adiabatic Growth of Massive Black Holes}

\begin{abstract}
We discuss the process of adiabatic growth of central black holes in the 
presence of a stationary, pre-existing distribution of collisionless stars.
Within the limitations of the assumptions, the resulting models make robust 
physical predictions for the presence of a central cusp in the stellar and 
dark matter density, a Keplerian rise in the velocity dispersion, and a 
significant tangential polarization of the velocity tensor. New generations of 
numerical models have confirmed and extended previous results, permit the 
study of axisymmetric and triaxial systems, and promise new insight into the 
dynamics of the central regions of galaxies.  These studies enable detailed 
comparisons with observations, further our understanding on the fueling 
processes for AGNs and quiescent black holes, and help elucidate the secular 
evolution of the inner regions and spheroids of galaxies. 
\end{abstract}
\section{Introduction}
Given the premise that the massive central dark objects in normal galaxies in
the local Universe are in fact supermassive black holes (Lynden-Bell 1969; 
Rees 1990; Kormendy \& Richstone 1995; Richstone et al. 1998), we can 
entertain a number of conjectures about the interaction of the central black 
hole with its environment. Obvious questions to consider include: formation 
scenarios for the black hole (e.g., Rees 1984; Shapiro in this volume); the 
demographics of the present population of black holes (Richstone and 
Ho in this volume); the fueling of active nuclei (Blandford in this volume);
the interaction of the active nucleus with its environment (Begelman in this 
volume); and, the effect of the central object upon the surrounding stellar 
population and the larger-scale structure of the host galaxy (Burkert, 
Gebhardt, Haehnelt, and Merritt in this volume). Hence, one can also test 
whether the inferred effects of the central object are consistent with 
observations, and whether additional observational constraints can be placed
on either the presence or the evolutionary history of the central black hole.

A particular assumption can be made (with the caveat, that, as with all 
assumptions, it may be false) that a substantial increase in mass of the 
central black hole takes place after initial formation, and that the mass is 
in some rigorous sense (to be established) added slowly to the pre-existing
seed black hole. This is the assumption of {\it adiabatic growth}, which will 
be reviewed here.  It leads to some nontrivial, testable predictions for the 
effects of black holes on their environments.

A central supermassive black hole dynamically dominates the surrounding 
stellar population inside some characteristic radius, $r_h = 
GM_{\rm BH}/\sigma^2$, where $M_{\rm BH}$ is the mass of the black hole and 
$\sigma$ is the velocity dispersion of the stars outside the 
radius of influence.  A natural ``shortest'' timescale for growth of a black 
hole is the ``Salpeter'' timescale, $t_S = M_{\rm BH}/\dot M_{\rm Edd} \approx 
5\times 10^{7} \ {\rm yr}$, where $\dot M_{\rm Edd}$ is the usual Eddington 
accretion timescale. The dynamical timescale inside $r_h$ is just $t_{dyn} = 
r_h/\sigma$.  For the Milky Way, $t_{dyn}(r_h) \sim 10^4 \ {\rm yr}$, for 
$M_{\rm BH} \sim 2\times 10^6 \,\msun $ and $\sigma \approx 66\, \kms$. If the 
observed correlation between dispersion and black hole mass holds, then 
$M_{\rm BH} \propto \sigma^4$ (Gebhardt \etal 2000a; Ferrarese \& Merritt 2000; 
Tremaine \etal 2002), and hence $t_{dyn} \propto \sigma $.  Hence we conclude 
that for reasonable black hole masses ($M_{\rm BH} \ltorder 10^{10} \,\msun$), 
the dynamical timescale inside $r_h$ is always much shorter than the Salpeter 
timescale, and therefore the likely timescale for black hole growth through 
accretion of baryonic matter is much longer than the dynamical timescale inside
the radius at which the black hole dominates the dynamics. We thus conclude 
that there may be a broad range of situations under which black hole growth is 
``adiabatic'' and the assumptions of these studies hold.  The stellar 
population will generally form a {\it density cusp}, $\rho \propto r^{-A}$, 
inside $r_h$, with the stellar velocity dispersion showing a Keplerian rise 
$\sigma (r) \propto r^{-1/2}$ inside the cusp (Peebles 1972; Bahcall \& Wolf 
1976; Young 1980; Quinlan, Hernquist, \& Sigurdsson 1995).

\subsection{Assumptions}
We consider the response of a stellar distribution function (\df) to the slow 
growth of a massive central black hole. The initial conditions assume there is 
no central black hole to begin with (or more realistically a seed black hole 
with a negligible initial mass), and there are usually implicit assumptions 
that the mass is ``magically'' added to the black hole --- that is to say, the 
mass of the central object is increased without necessarily withdrawing the 
mass from some explicit reservoir.  The underlying assumption here is that the 
mass is accreting from some diffuse medium, like cool gas, that is distributed 
like the stellar population but with a density much lower than that of the 
stellar mass density, and is replaced by some inflow that is an implicit outer 
boundary condition (e.g., Young 1980; Quinlan \etal 1995).  This is {\it not}\
a necessary assumption; it is just a simplifying assumption. It is trivial to 
extend it to scenarios where the mass is explicitly withdrawn from some 
reservoir, with the added complication of having to specify the physical 
nature of the mass reservoir. In most situations modeled so far, where the 
mass comes from is not important; the response of the system is robust, 
independent of the source of the mass. The exception is if all the mass comes 
from the black hole swallowing the most tightly bound stars only, in which 
case the conclusions are somewhat different and the process effectively 
violates our assumptions of adiabatic growth.

An additional implicit assumption is that the black hole is {\it central}; 
that is, it is at the center of mass of the stellar system. In practice, the 
surrounding stellar system is discrete and the black hole mass is finite, so 
we expect the black hole to undergo quasi-Brownian motion away from the center 
(see, e.g., Chatterjee, Hernquist, \& Loeb 2001). For masses of astrophysical 
interest, the displacement is typically much larger than the black hole 
Schwarzschild radius, $r_S$, but much smaller than $r_h$, and the timescale 
for wandering is short enough that the outer cusp is not carried with the 
black hole as it moves; this can lead to modification of the cusp profile at 
small radii ($r \ltorder 10^{-3} r_h$ for typical $M_{\rm BH}$) as the black 
hole wandering produces rapid fluctuations in the central potential seen by 
stars in the inner cusp (see Sigurdsson 2003). 

A final fundamental consideration is whether the dynamics of the stellar 
population are ``collisionless'' --- that is, whether the relaxation timescale 
for a population of $N$ stars, $t_R \sim N t_{dyn}/8 \ln \Lambda$, is shorter 
or longer than the evolutionary timescale of the stellar system, usually taken 
to be the Hubble time, $t_H \sim 10^{10} \ {\rm yr}$ (e.g., Spitzer 1971; 
Hills 1975). The response of a relaxed stellar system to the presence of a 
central massive black hole has been extensively considered, primarily in the 
context of globular clusters, or in the context of initial black hole 
formation and rapid growth in protogalaxies  (Bahcall \& Wolf 1976, 1977; 
Lightman \& Shapiro 1977; Cohn \& Kulsrud 1978; Shapiro \& Marchant 1978; 
Shapiro 1985; Amaro-Seoane \& Spurzem 2001 and Freitag \& Benz 2001 and 
citations therein).  For supermassive black holes in normal, evolved, galaxies
the relaxation timescales in the inner spheroid, but outside the black hole 
cusp, are generally longer than the Hubble time; inside the cusp the 
relaxation time may be constant, increase, or decrease with decreasing radius. 
For those cases where the relaxation time decreases with decreasing radius, 
the dynamics of the stellar population surrounding the black hole {\it may}\
undergo a transition to the fully collisional regime in the inner cusp, and 
the discussion in the papers cited above then becomes appropriate but is beyond
the scope of this review. The mean central relaxation time can be approximated 
as $t_R \sim 2\times 10^9 (\sigma/200 \, {\rm km\, s^{-1}})^3/(\rho /10^6 \, 
\msun {\rm pc^{-3}})$ (Young 1980). It is not sufficient that $t_R < t_H$ for 
non-adiabatic growth.  For such relaxed cusps the relaxation time at small 
radii may become shorter, and, if $t_R \ltorder t_S$ at some small radius, 
which may well occur for a significant fraction of galactic nuclei or 
proto-nuclei at some point in their evolution, then any central black hole may 
grow by tidal disruption of stars or by swallowing stars whole, more rapidly 
then the cusp can dynamically readjust its structure; in such a situation, 
the growth is definitely non-adiabatic.

The underlying physical assumption of the ``adiabatic growth'' model is that 
as the integrals of motion change smoothly in response to the increase in 
central mass, the action variables for the surrounding stellar population 
remain invariant (Binney \& Tremaine 1987).  This is to be contrasted with the 
opposite extreme assumption of ``violent relaxation,'' in which the potential 
is assumed to fluctuate rapidly compared to the dynamical time, and the \df
evolves to some final statistical equilibrium state (Lynden-Bell 1967; 
Stiavelli 1998). 
The resulting ``final distribution'' may then be compared with observations. 
It should be noted that real galaxies may not have ``initial'' DFs that are 
well represented by any of the analytic or numerical distributions assumed in 
these models, nor is it necessarily the case that significant increase in 
black hole mass ever takes place under conditions in which the adiabatic 
approximation holds. In particular, an implicit assumption is that a relaxed 
stellar population is in place as an initial condition, and that significant 
increase in black hole mass takes place {\it after}\ (the inner region of) the 
galaxy is assembled.  The adiabatic models are physically distinct from 
{\it ab initio}\ models, where a \df including a central black hole, by design,
is required to satisfy the Boltzmann equation (e.g., Huntley \& Saslaw 1975;
Tremaine \etal 1994).  The adiabatic models are also distinct from the  ``orbit assembly'' models used to construct kinematic models of observed galaxies
(Schwarzschild 1979; Richstone \& Tremaine 1984, 1988; Magorrian \etal 1998). 

An interesting question is whether any of these models in some sense 
rigorously represent real stellar systems. Nature need not settle on the 
analytically or numerically derived solutions of the Boltzmann equation, out 
of the infinite number that exist.  As found by Quinlan \etal (1995), 
apparently small differences in some phase-space values can lead to large 
changes in the averaged properties of the evolved system.  We may also worry 
whether the different techniques for constructing stationary solutions of the 
Boltzmann equation representing collisionless stellar objects surrounding a 
central black hole are actually equivalent, or whether the different 
techniques produce wholly distinct families of solutions, as opposed to 
solutions with an overlap in properties or formally identical for some range 
of parameters.

\section{Spherical Growth}

The response of a spherical distribution of stars to the adiabatic growth of a 
central black hole in the collisionless limit was first considered by Peebles 
(1972) for an isothermal sphere.  Young (1980) confirmed the primary result 
that a density cusp $\rho \propto r^{-3/2}$ would form, with an associated 
velocity dispersion cusp, $\sigma (r) \propto r^{-1/2}$; he also pointed out 
that the velocity anisotropy, 
%XX factor of 2 difference in denominator??
$\beta(r) = 1 - \langle v_t^2\rangle/\langle 2v_r^2\rangle $, becomes 
negative (tangentially biased) at small radii, where $v_t$ and $v_r$ is the 
tangential and radial velocity, respectively.  Goodman \& Binney (1984) showed 
that when $\beta (0) = 0$ the distribution is isotropic at the center for an 
initial isothermal distribution (see also Binney \& Petit 1989), and Lee \& 
Goodman (1989) generalised the approximate solution of the problem to 
axisymmetric rotating distributions.  The basic physics of the problem for 
a spherical system are discussed in Shapiro \& Teukolsky (1983 and reference 
therein), as a simple application of Liouville's theorem.  Their Equation 
14.2.9 shows the response of a spherical system, with some initial DF $f(E)$, 
to a central black hole.  The final density 
$n(R) = 4\pi \int f(E) \sqrt{[2(E - \Phi)]} dE \propto r^{-1/2}\times r^{-1}$ 
for $f(E) \rightarrow {\rm constant}$, appropriate for the $n=0$ case 
discussed by Quinlan et al. (1995). 

Quinlan \etal (1995) generalised the result to a broad range of initially 
spherical DFs and found that for different \df the final cusp slope may be 
very different, even for near-identical initial spatial density profiles. They 
also found that, in contrast with the result for initially isothermal 
distributions, for some initial \df the polarization of the velocity 
distribution is generic and always tangentially biased, and that the 
tangential bias may persist to zero radius.  The velocity distribution is in 
general non-Gaussian, and initially non-Gaussian distributions may evolve
to be either closer to or farther from Gaussian in response to the black hole 
growth (Sigurdsson, Hernquist, \& Quinlan 1995).  The net results are 
distinct, but unfortunately not provide a simple or unique prediction for the 
final spherical distribution of a stellar population responding adiabatically 
to the growth of a central black hole.  The semi-analytic results of Quinlan 
\etal were confirmed numerically in a companion paper by Sigurdsson \etal 
(1995), who extended the numerical methodology to a family of non-spherical 
models.

A major purpose for producing a broad range of adiabatic growth models is for 
comparison with observations, for example to establish robust estimators for 
central black hole masses from the observed surface density profiles or
spectroscopically determined projected velocity dispersion profiles.  In 
addition to the intrinsic degeneracies between the \df and the density and 
dispersion profiles, we are mostly restricted  to observing projected 
quantities, the line integrals of the light density and velocity 
distribution, which lead to degeneracies in the inversion to the full volume 
distribution (e.g., Romanowsky \& Kochanek 1997 and references therein).
We are further restricted to observing the dominant light-emitting population 
(mainly giant, sub-giant and post-AGB stars), and the mass may be distributed 
differently, with different stellar populations (or dark matter) having 
different density profiles.  Still, with the use of higher moments of the 
velocity distribution (van der Marel \& Franx 1993; Dehnen \& Gerhard 1994;
van der Marel 1994a, b; van der Marel \etal 1994) strong constraints can be 
put on the true stellar \df; by making some ``natural'' assumptions (e.g., 
the unobserved dark matter distribution is consistent with the light 
distribution), strong constraints can be put on the total mass of any inferred 
central dark object.

\subsection{Action}

In a spherical potential, we consider some initial \df $f$ specified by the 
energy $E$ and angular momentum $L$.  The quantity $f$ is then also a function 
of the actions $L$ and $J_r = \oint v_r dr$. As the integrals change under the 
adiabatic growth of the black hole, the action, by assumption, remains 
invariant, and $f$ evolves to remain a fixed function of the actions (see 
Young 1980 and Quinlan \etal 1995 for discussion).

We want to consider some initial \df with an explicitly specified form and a 
corresponding density profile (see Binney \& Mamon 1982; Binney \& Tremaine 
1987).  Of particular interest is the asymptotic behaviour of the density 
profile at small radii ($\rho(r) \propto r^{-\gamma}$; as $r\rightarrow 0$)
and the corresponding asymptotic behaviour of $f(E)$ in the limit 
$E\rightarrow \Phi (0)$, which in general is some power law 
$f(E) \sim [E - \Phi (0)]^{-n}$.  (But note that real galaxies need not be 
nicely monotonic power laws, even asymptotically, but may, for example, have 
density inversions at small radii (e.g., Peebles 1972; Lauer \etal 2002.) For 
an isothermal density profile the central density approaches a constant at 
small radii, as does the \df at the lowest energies. In general, 
$0 \leq \gamma \leq 3$, and it is useful to distinguish between models with 
``analytic'' cores (in the nomenclature of Quinlan \etal 1995), which have a 
density profile $\rho(r) \approx \rho_0 + {1\over 2} \rho'' r^2 + \ldots$, as
$r\rightarrow 0$, and non-analytic models, which do not approximate a harmonic 
potential at the origin. Particular examples of analytic models include a
non-singular isothermal sphere, a King model, a Plummer model, or an isochrone.
Non-analytic models include (1) a singular isothermal sphere (Cipollina \& 
Bertin 1994); (2) the $\gamma = 2$ Jaffe (1983) or $\gamma = 1$ Hernquist 
(1990) model; (3) the generalised ``gamma'' models ($0 \leq \gamma \leq 3$; 
Dehnen 1993; Tremaine \etal 1994); (4) and further spherical generalisations 
of these models, such as those of Navarro, Frenk, \& White (1997) and Zhao 
(1997).

It turns out that the final density cusp slope $A$ generally depends both on 
the initial density slopes $\gamma$ and the asymptotic divergence of the \df 
$n$; the relationship among the variables is given analytically 
by\footnote{For the full derivation, see Appendix A of Quinlan \etal (1995), 
Gondolo \& Silk (1999), and Ullio, Zhao, \& Kamionkowski (2001).}
 
\begin{equation}
A = {3\over 2} + n\left ( {{ 2 - \gamma}\over {4 - \gamma}} \right ).
\end{equation}
 
As illustrated in Figure 1, which compares the adiabatic growth of a black 
hole in a $\gamma = 0$ model with an isochrone model (H\'enon 1960), analytic 
and non-analytic models with the same, or very nearly the same, initial 
density profiles produce qualitatively different final density profiles.
\begin{figure}
\plottwo{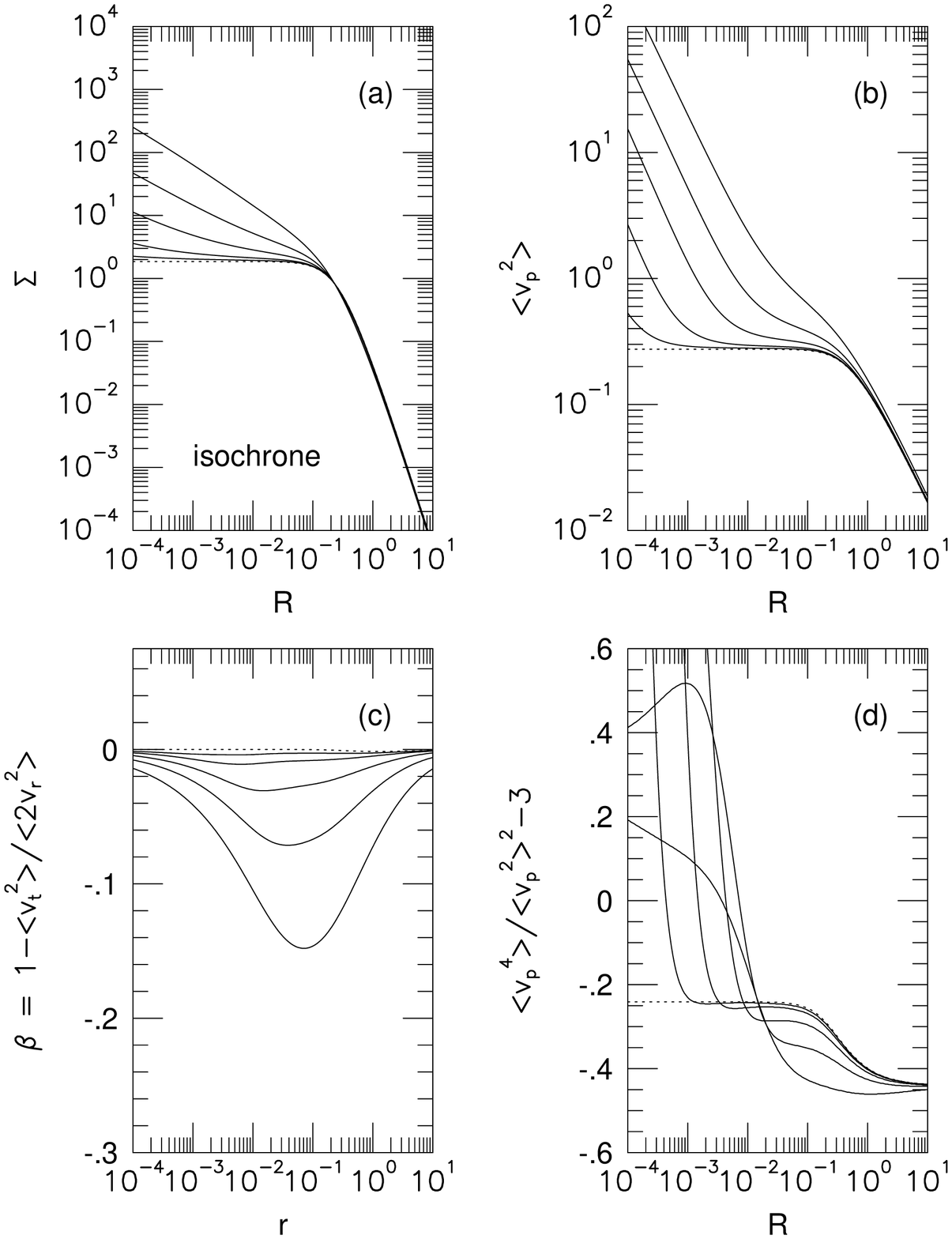}{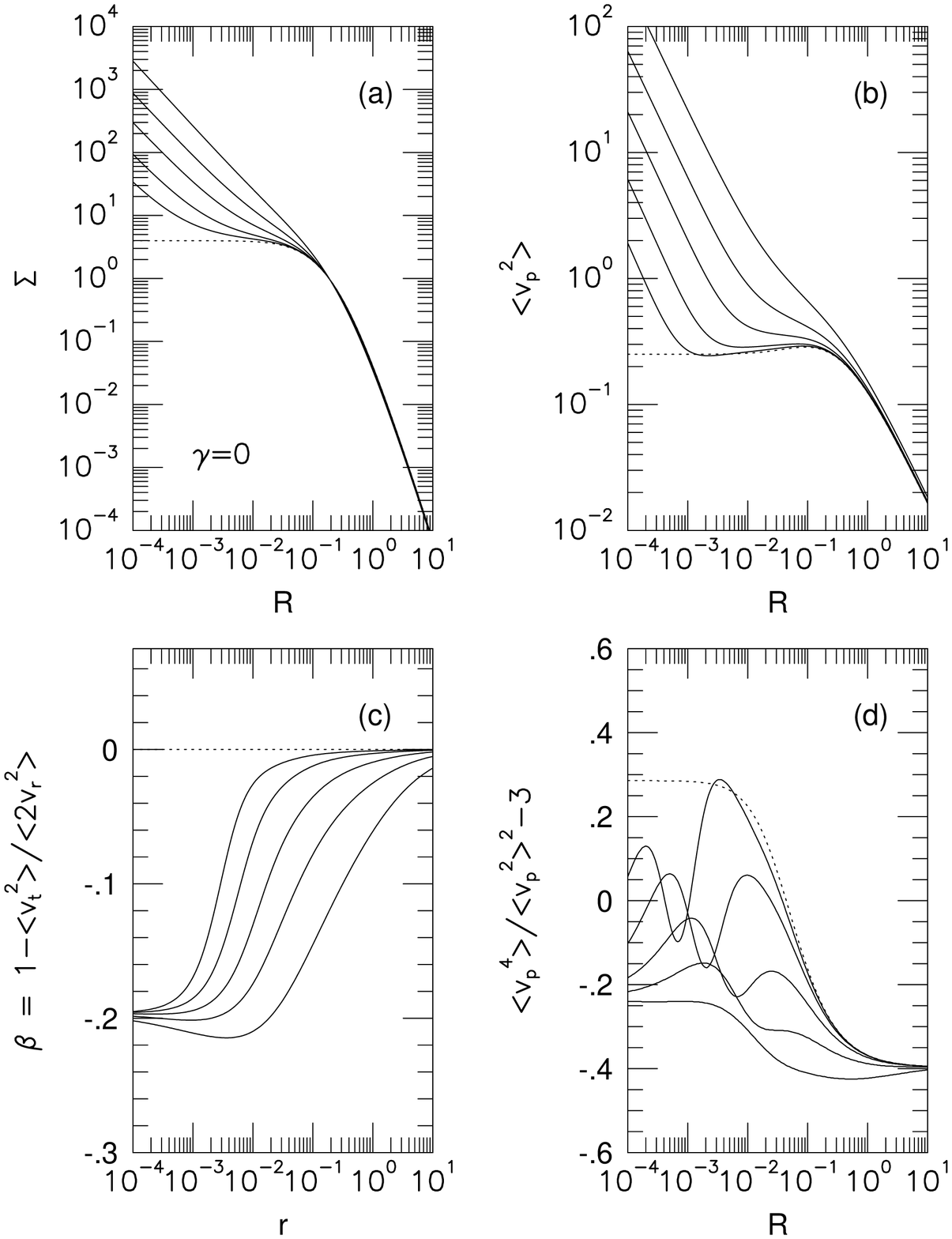}
\caption{Results of the growth of black holes of different masses, 
$M_{\rm BH} = 10^{-3} - 10^{-1}$ of the total (spheroid) galaxy mass, in an 
isochrone and $\gamma = 0$ model, respectively.  The dotted lines show the 
initial (near-identical) models.  The panels show ({\it a}) surface density, 
({\it b}) projected velocity dispersion, ({\it c}) projected anisotropy 
$\beta (R)$, and ({\it d}) kurtosis $\kappa -3$. Note the very different final 
profiles despite near-identical observable initial conditions. (From Quinlan 
\etal 1995.)}
\label{ss-figure1}
\end{figure}
More generally, the response to the adiabatic growth of a black hole produces 
a cusp with slope as low as $A = 3/2$, as originally found, up to values as 
steep as $A=3$, although $A=5/2$ is probably the steepest physically 
sustainable slope before collisional effects in the inner cusp necessarily 
dominate the dynamics.  Table 1.1 lists the values of some initial and final 
slopes (Quinlan \etal 1995).  $C$ is the final cusp slope in the limit of an 
initially completely tangentially biased \df.
 \begin{table}
  \caption{Adiabatic density cusps}
    \begin{tabular}{l|cccc}
     \hline \hline
     {Model} & {$\gamma$} & {$n$} & {$A$} & $C$ \\
     \hline
     isochrone & $0$ & $0$ & $3/2$ & $9/4$ \\ 
     $\gamma = 0$ & $0$ & $1$ & $2$ & $9/4$ \\ 
     $\gamma = 1$ & $1$ & $5/2$ & $7/3$ & $7/3$ \\ 
     $\gamma = 3/2$ & $3/2$ & $9/2$ & $12/5$ & $12/5$ \\ 
     $\gamma = 2$ & $2$ & $-$ & $5/2$ & $5/2$ \\ 
     \hline \hline
    \end{tabular}
  \label{ss-table1}
 \end{table}
Note that the presence of a density cusp by itself is {\it not}\ a robust 
indicator of a central supermassive black hole; this is clearly so, since, for 
example, the Jaffe (1983) model or singular isothermal sphere, with no central 
black holes, have $\gamma = 2$ cusps, steeper than the $A=3/2$ cusps predicted 
for the response of a non-singular isothermal sphere to a central black hole.
On the other hand, the presence of a density cusp, a Keplerian rise in the
velocity dispersion, and the kinematic signatures of tangential anisotropy at 
small radii {\it are}\ robust indicators of a central 
supermassive black hole.

This does not preclude the possibility that in the absence of a central black 
hole actual stellar systems tend toward flat, constant density cores, whether 
through formation or relaxation, and that in practice cusps are in fact 
signatures of central black holes. We know that for a broad range of formation 
scenarios, stellar cusps form around central black holes (with cusps as 
shallow as $A=1/2$ or as steep as $A=5/2$); however, it is possible that in 
some situations binary black holes completely destroy cusps, leaving density 
inversions (e.g., Peebles 1972), and we know it is possible for cuspy stellar 
systems to exist in the absence of central black holes.  Assuming that cusps 
are tracers of black holes, van der Marel (1999) has explored the use of the 
adiabatic growth models in matching observed density profiles.

\subsection{Anisotropy}

Quinlan \etal (1995) also experimented with spherical, radially anisotropic 
distributions (Osipkov 1979; Tonry 1983; Merritt 1985; Dejonghe 1987; 
Cudderford 1991; Gerhard 1993), but found it impossible to generate physical 
distributions with significant radial anisotropy persisting to zero radius. 
The general conclusion is therefore that adiabatic growth induces tangential 
bias at small radii, and that an initial tangential bias can, but does not 
necessarily, lead to steeper final cusps, compared to the equivalent isotropic 
model. The Keplerian velocity cusp is a robust prediction of spherical 
adiabatic growth models. As noted by Duncan \& Wheeler (1980), however, a 
strong radial velocity anisotropy can mimic a Keplerian rise in velocity in 
projection, although there are severe concerns about the stability of any such 
models (Merritt 1987; Palmer \& Papaloizou 1988). A robust prediction of a 
tangential bias induced by any central black hole is therefore potentially 
important, although the anisotropy is not a directly observable quantity but 
must be inferred from the projected moments of the velocity distribution.

As shown in Figure 1.1, the final anisotropy, $\beta (r)$, may be either zero 
at the black hole or remain negative at small radii. The deviation from 
Gaussianity is conveniently measured by the kurtosis, $\kappa$ (by construction,
the skew is zero for these models), or equivalently, the fourth Gauss-Hermite 
moment, $h_4 \approx (\kappa - 3)/8\sqrt{6}$, for $h_4 \ltorder 0.03$ 
(van der Marel \& Franx 1993; Dehnen \& Gerhard 1994; Quinlan \etal 1995).

With $A = 3/2 + p > 3/2$, the relaxation timescale at small radii decreases as 
$t_R \propto r^p$.  The cusps induced in isotropic, analytic models have $p=0$ 
and constant $t_R$; more generally, $p > 0$ and $t_R$ can be small close to 
the black hole.  Very close to the black hole, relaxation and collision 
timescales get short for strong cusps, and strong collisional effects may lead 
to rapid growth of the black hole, with corresponding associated depletion of 
the stellar population.  This is certainly the case for cusps as steep as 
$A = 3$, and may even be a problem for shallower cusps (Frank \& Rees 1975; 
Quinlan \& Shapiro 1990; Quinlan \etal 1995; Sigurdsson \& Rees 1997; Freitag 
\& Benz 2001).  

\subsection{Non-adiabatic growth}

Formally, the adiabatic growth model implies an infinitely long timescale
for accretion. In practice, of course, any growth in mass occurs on a finite 
timescale.  We can investigate the nature of non-adiabatic growth without 
losing the predictive power of the adiabatic models.

Sigurdsson \etal (1995; see also Hernquist \& Ostriker 1992; Hernquist, 
Sigurdsson, \& Bryan 1995; Sigurdsson \etal 1997a) explored the timescale for 
adding mass, and concluded that, for timescales $t \gtorder 10\, t_{dyn}(r_h)$, 
the adiabatic approximation was satisfied for the resolution of the models. 
The use of $N$-body modeling also showed the final distributions after 
adiabatic growth was stable; stability is not guaranteed by the adiabatic 
growth process, nor is there a general analytic criterion for stability of 
arbitrary \df. 

Adiabatic growth formally also implies reversibility. Sigurdsson \& Hernquist 
(unpublished) experimented with numerical models in which a central black hole 
grown adiabatically in a spherical stellar distribution was {\it removed}\ 
adiabatically. The original distribution was in fact recovered to within the 
resolution of the models.

In general, violent formation can lead to either galaxies with constant-density
cores (e.g., Lynden-Bell 1967; van Albada 1982; Norman, May, \& van Albada 
1985; Burkert, this volume) or singular profiles (e.g., Aarseth 1966; Fillmore 
\& Goldreich 1984; Bertschinger 1985; Navarro et al. 1997). 

Stiavelli (1998) and Ullio \etal (2001) explored non-adiabatic growth with 
a pre-existing black hole and found results that did not deviate strongly from 
the case of adiabatic growth.  More recently, MacMillan \& Henriksen (2002) 
suggested that non-adiabatic accretion of dark matter might account for the 
$M_{\rm BH}-\sigma$ relation, which is not explained by a simple adiabatic 
compression of the dark matter halo (Dubinski \& Carlberg 1991).  Adiabatic 
compression of the dark matter in the inner regions by the formation of a 
central black hole is potentially interesting, as it can lead to increased 
rates of dark matter accretion onto the black hole, and to higher rates of 
dark matter self-interaction, for models in which such interactions may occur 
(Gondolo \& Silk 1999; Ostriker 2000).

Sigurdsson \etal (1995) also found that steep initial density cusps were 
vulnerable to violent disruption by the ``wandering'' of the central black 
hole.  Black hole mergers will also efficiently destroy steep stellar cusps 
around a black hole (Makino \& Ebisuzaki 1996; Quinlan \& Hernquist 1997; 
Faber \etal 1997; Milosavljevi\'c \& Merritt 2001; Zier \& Biermann 2001; 
Hemsendorf, Sigurdsson, \& Spurzem 2002; Ravindranath, Ho, \& Filippenko 
2002). 

\section{Non-spherical Systems}

The obvious next approximation beyond spherical (isotropic and anisotropic) 
models is to consider axisymmetric ones. A number of families of two- and 
three-integral axisymmetric models exist in the literature (e.g., Evans 1993; 
Hunter \& Qian 1993; Kuijken \& Dubinski 1994; Qian \etal 1995; Gebhardt \etal 
2000b; Lynden-Bell 2002).

Van der Marel \etal (1997a) constructed a detailed model for the central black 
hole in M32, and van der Marel, Sigurdsson, \& Hernquist (1997b) ran a 
numerical model, using techniques developed for adiabatic growth simulations, 
to demonstrate its stability.  A concern remains that such models may be 
unstable to $m=1$ modes, which are typically suppressed in numerical 
simulations (if not, they can arise spontaneously through numerical artifacts, 
which make it difficult in general to identify physical instabilities).  Lee 
\& Goodman (1989) modeled adiabatic growth of black holes in approximate 
rotating, isothermal axisymmetric models.  Rather interestingly,  they found 
that the rotation curve rises more rapidly than the dispersion curve, but not 
enough to account for the observed high rotation in the inner regions of some 
systems.  A more general exploration of adiabatic growth in non-rotating 
axisymmetric models was done by Sigurdsson \& Hernquist (unpublished, see 
Fig. 2). They found that axisymmetric models are quite similar to spherical 
models, particularly in that the tangential anisotropy is induced in the cusp 
that the density profile becomes rounder at small radii (Fig. 2).  Leeuwin \& 
Athanassoula (2000) simulated adiabatic growth in Lynden-Bell (1962) models 
and obtained results consistent with those of Lee \& Goodman (1989), including 
rounding of the inner density profile and a significant rise in the rotation 
velocity inside the cusp, consistent with the tangential polarization of the
central black hole. 
\begin{figure}
\plottwo{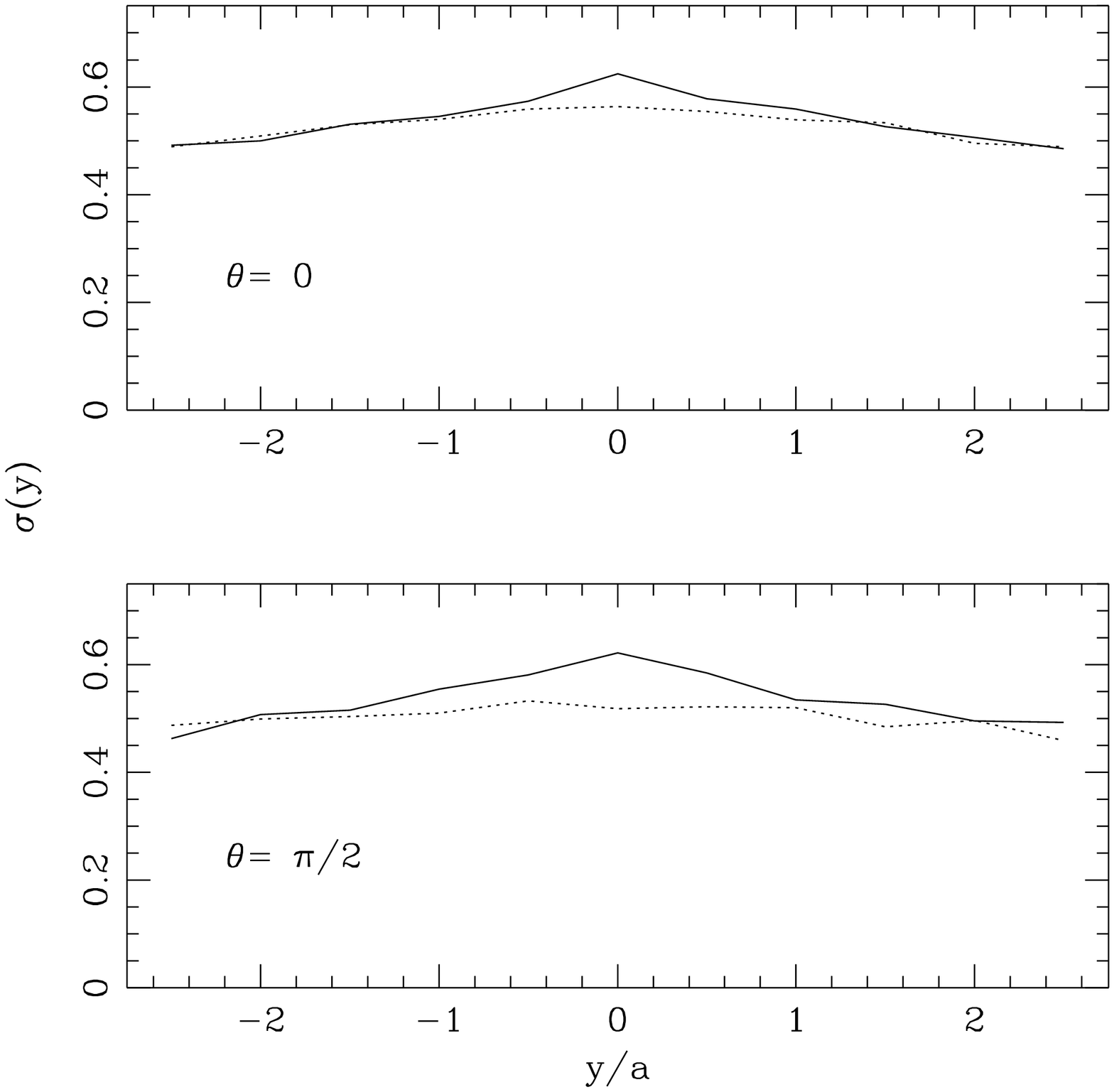}{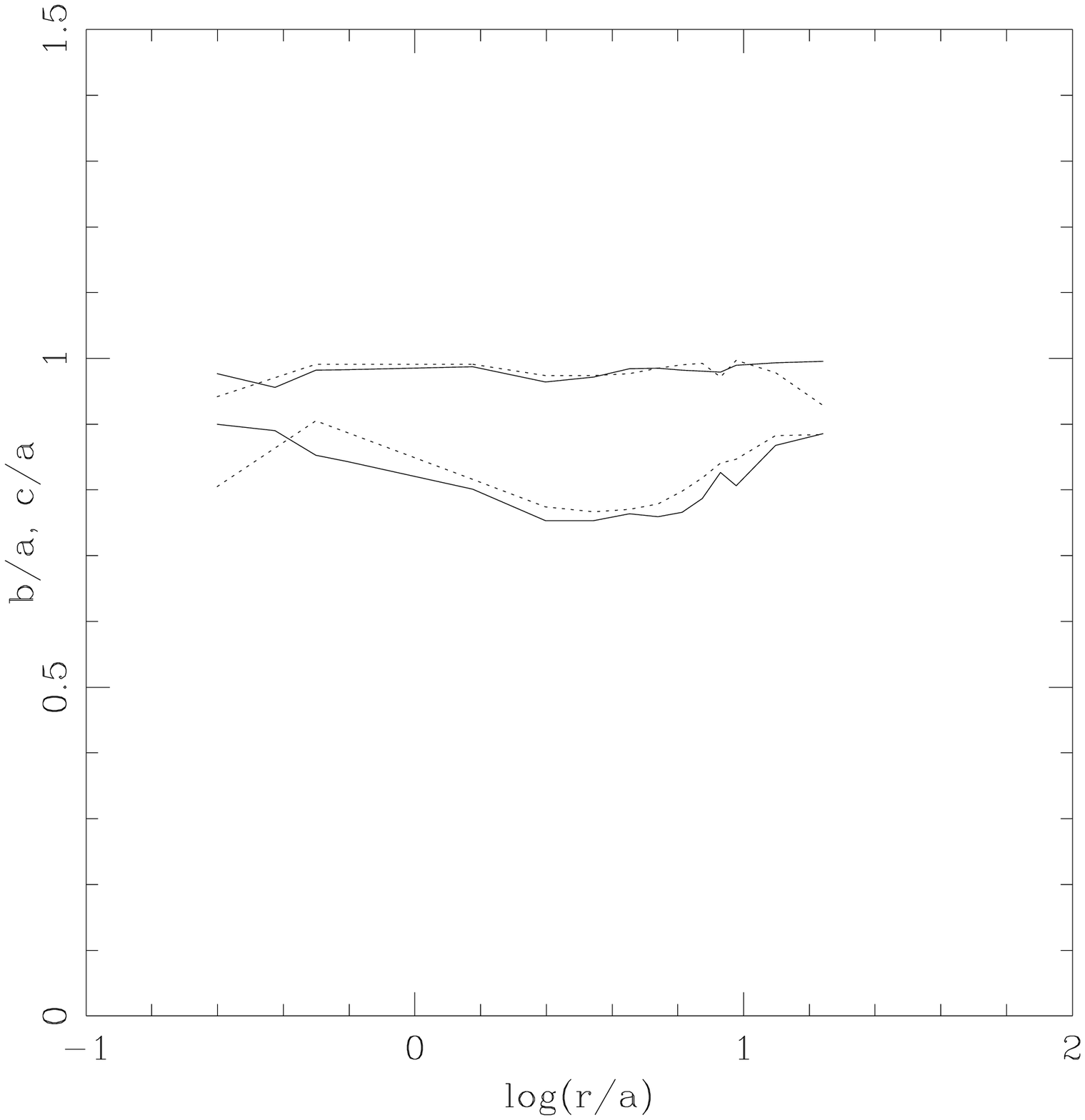}
\caption{Results of the growth of a black hole in an ``Evans model.'' The 
figures show the projected density and axis ratios for edge-on and face-on 
inclinations before ({\it dotted}\ lines) and after ({\it solid}\ lines) lines 
black hole growth.  The inner density profile becomes rounder in response to 
the black hole, and the characteristic Keplerian dispersion profile is seen in 
projection. (From Sigurdsson \& Hernquist, unpublished.)}
\label{ss-figure2}
\end{figure}
\subsection{Triaxial systems}
Real galaxies are generally not spherical or axisymmetric, but triaxial 
(Binney 1976; Franx, Illingworth, \& de Zeeuw 1991; Ryden 1992, 1996; Tremblay 
\& Merritt 1995; Bak \& Statler 2000).  We expect triaxial galaxies to form 
from general cosmological initial conditions (e.g., Norman \etal 1985; 
Dubinski \& Carlberg 1991) and from galaxy mergers (Barnes 1988, 1992; 
Hernquist 1992, 1993).  Exact analytic models exist for triaxial galaxies with 
cores (Schwarzschild 1979; de Zeeuw 1985; Statler 1987; van de Ven \etal 2002). 
Observationally, we also see that the density profiles of the spheroidal 
component of galaxies generally continue to rise toward the center, with 
$0.5 \ltorder \gamma \ltorder 2.3$ (e.g., Lauer \etal 1995; Gebhardt \etal
1996; Faber \etal 1997; Ravindranath et al. 2001).  There are also observed 
correlations between the cusp slope $\gamma$ and the global properties of the 
galaxy, including shape in the form of boxy or disky isophotes (e.g., Faber 
\etal 1997). 

Historically, dynamical arguments suggest that the presence of a strong 
central cusp ($\gamma > 1$) induces chaos in the orbit families thatb populate 
the galaxy, driving the system away from strong triaxiality (e.g., Gerhard \& 
Binney 1985; Norman \etal 1985; Merritt \& Valluri 1996; Merritt \& Quinlan 
1998; Merritt 1997, 1999).  The argument is that central cusps or central 
point masses scatter the box orbits that support triaxiality in galaxies, 
inducing a population of chaotic orbits which drive figure evolution toward 
axisymmetry (Miralda-Escude \& Schwarzschild 1989; Lees \& Schwarzschild 1992; 
Fridman \& Merritt 1997; Valluri \& Merritt 1998).  Hence, central 
supermassive black holes should preclude the presence of triaxiality at small 
radii, and might drive global figure evolution of the system (Norman \etal 
1985; Merritt \& Quinlan 1998).  This is potentially very important because 
triaxial potentials support fueling of the central black hole through material 
falling into it on box orbits (e.g., Norman \& Silk 1983; Valluri \& Merritt 
1998) or by gas traveling on intersecting orbits that drive dissipation and 
inflow, thus providing a direct link between the dynamics in the center of the 
galaxy and its global properties.  In the extreme case of disk systems, 
analogous instabilities exist (e.g., Hasan \& Norman 1990; Sellwood \& Valluri 
1997).

\subsection{Adiabatic growth and triaxiality}

Holley-Bockelmann \etal (2001; see also Sigurdsson \etal 1997b, 1998) showed 
that applying numerical adiabatic growth techniques to ``squeeze'' an initially 
spherically symmetric cuspy DF could produce a stable, stationary, cuspy 
triaxial configuration with well-characterised phase-space properties. A key 
aspect of the models is that they contain a central cusp of near-constant 
slope and near-constant axis ratios with significant triaxiality at all radii
resolved by the models (Holley-Bockelmann et al. 2001, 2002).  Galaxies with 
density cusps support different stellar orbits than, for instance, $\gamma=0$ 
core models (Gerhard \& Binney 1985; Gerhard 1986; Pfenniger \& de Zeeuw 1989; 
Schwarzschild 1993; de Zeeuw 1995; Merritt 1999; Holley-Bockelmann et al. 2001).
The set of models thus produced provide a starting point for investigation of 
the adiabatic growth of central black holes in cuspy, triaxial potentials. A 
black hole is then grown using the previously developed numerical $N$-body 
techniques (Sigurdsson \etal 1995; Holley-Bockelmann \etal 2002).

Following Holley-Bockelmann \etal (2001, 2002), consider a black hole grown in 
a triaxial Hernquist model with initial cusp slope $\gamma = 1$.  As the black 
hole grows, both the cusp slope $\gamma$ and central velocity dispersion 
$\sigma_p$ increase, as in spherical and axisymmetric models.  The cusp 
settles to an equilibrium value $\gamma \simeq 2.05$, measured at projected 
ellipsoidal radius $Q = 10^{-1.3}$, with projected central dispersion 
$\sigma_p \simeq 0.7$, measured at projected ellipsoidal radius $Q = 10^{-2.3}$.
These results are characteristic of adiabatic black hole growth in cuspy 
galaxies and can be compared both to analytic estimates for adiabatic black 
hole growth in a spherical $\gamma=1.0$ model, which predict $\gamma = 7/3$ 
and $\sigma_p=0.75$ (Quinlan \etal 1995), and to the results from $N$-body 
simulations where $\gamma \approx 2.2$ and $\sigma \approx 0.65$ (Sigurdsson 
\etal 1995).  The fact that the measured cusp slope is less than the analytic 
value is to be expected, since the cusp slope is measured over a finite radial 
range near the center, and it is not the asymptotic $q=0$ value.  

As the black hole grows, the inner regions become rounder (Fig. 3{\it b}); 
the central 10\% of the mass, corresponding to an ellipsoidal radius 
$q$ $=\sqrt{ x^2 + (y/b)^2 + (z/c)^2}< 0.1$, is close to spherical with axis 
ratios $a:b:c= 1.0:0.95:0.92$.  The shape evolution in the outer regions is 
much less dramatic.  Following the growth of the black hole, the model 
exhibits a marked shape gradient, becoming more strongly triaxial with 
increasing radius.  Despite the nearly axisymmetric shape at the center, the 
inner region is still triaxial enough to influence the stellar-orbital dynamics
(Statler 1987; Hunter \& de Zeeuw 1992; Arnold, de Zeeuw, \& Hunter 1994).
 
\begin{figure}
%\plottwo{ssfig3.ps}{ssfig4.ps}
\plotone{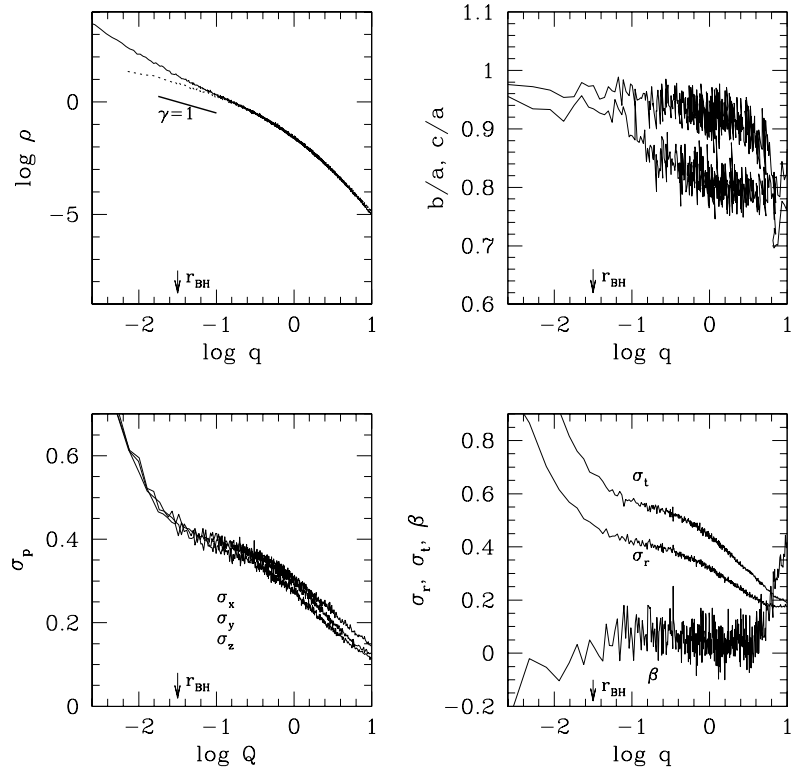}
\caption{The structural and kinematic properties of a cuspy triaxial 
galaxy model with a massive central black hole.  {\it Upper left}: density 
profile; {\it upper right}: intermediate and minor axis lengths as a function 
of ellipsoidal radius; {\it lower left}: projected velocity dispersion
along the fundamental axes, as a function of projected ellipsoidal radius; 
{\it lower right}: true radial and tangential velocity dispersion, and 
velocity anisotropy parameter, as a function of ellipsoidal radius. (From 
Holley-Bockelmann \etal 2002.)
}
\label{ss-figure3}
\end{figure}
 
The final state of this model features several hallmarks of a black 
hole-embedded triaxial figure.  Figure 3 shows the properties of this object 
as a function of ellipsoidal radius $q$ at $T=40$ (12.8 $t_{\rm dyn}$ at 
$q=1$), well after the model black hole has stopped growing.  Figure 3{\it a}\
shows the $\gamma\approx 2$ density cusp induced by the black hole inside 
$\log q = -1$.  At a larger radii $\log q > -1$, however, this plot 
demonstrates that the system retains the original Hernquist density profile. 
Figure 3{\it b}\ shows explicitly the strong shape gradient in the model.  
Inside $r_h$, both the projected and intrinsic velocity dispersions 
exhibit a strong central cusp (panels {\it c}\ and {\it d}). In the outskirts, 
where the model maintains its triaxiality, the projected velocity 
distributions follow $\sigma_x>\sigma_y>\sigma_z$, in accord with a triaxial 
model where $a>b>c$. However, inside the cusp the projected velocity 
dispersions are commensurate. Interestingly, the anisotropy parameter, 
$\beta = 1-\langle v_t^2\rangle / \langle 2 v_r^2\rangle$, becomes negative near 
the black hole.  This is consistent with models of stellar orbits around a 
black hole that is adiabatically grown, where $\beta=-0.3$ (Goodman \& Binney 
1983; Quinlan et al. 1995). Exterior to the black hole's radius of influence, 
the system is radially anisotropic ($\beta>0$), as expected for a triaxial 
galaxy.

Poon \& Merritt (2001, 2002), using Schwazschild's orbital-assembly technique,
have now also found models with triaxiality at small radii in the presence of
a central black hole and a surrounding cusp.

Clearly, it is possible for some significant triaxiality to persist both in 
the presence of a central cusp, and, more importantly, in the presence of a 
central black hole.

\subsection{Chaos}

If a significant fraction of the orbits in triaxial models containing central
black holes become chaotic, then by the ergodic theorem the shape of the
distribution must evolve toward sphericity (possibly halting when axisymmetry
is reached).  It is clear that the onset of chaos in the most tightly bound 
orbit families leads to a rapid change in the inner structure of the model
galaxies. In the outer regions, orbits stay regular even after repeated 
passages near the potential center. It is possible that many of these orbits
are actually chaotic orbits that are ``sticky'' (Siopis \& Kandrup 2000), with 
a very long diffusion timescale.  The course grainedness of the numerical 
model potential seems to argue against this explanation; a course-grained 
potential effectively creates holes in the Arnold web (Arnold 1964) through 
which an otherwise confined orbit may escape. Figure 4 illustrates what is 
probably happening in these models; strong scattering is taking place, 
inducing some chaos, but triaxiality is sustained by the persistence of 
resonant boxlet orbits that scatter between each other, rather than into true 
chaotic orbits.

\begin{figure}
\plotone{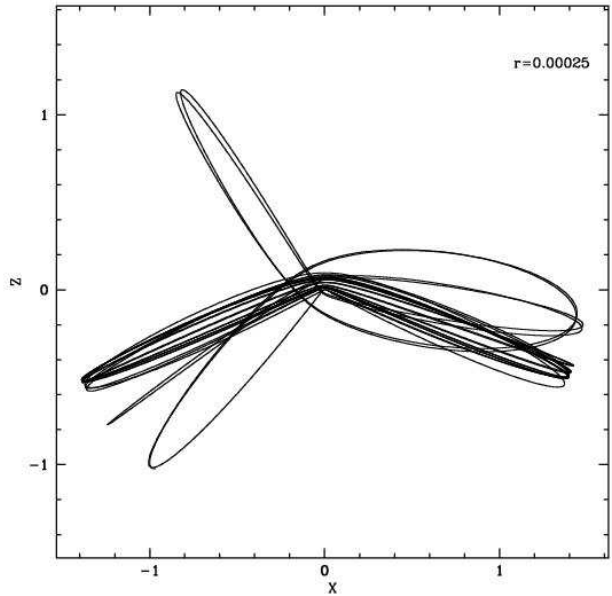}
%\plotfiddle{ssfig4.ps}{4in}{0}{100}{100}{0}{0}
\caption{The evolution of a banana boxlet 
orbit due to scattering after close approach to a central black hole in a 
triaxial potential. The banana orbit flips to a fish orbit, another resonant 
boxlet, and does not become chaotic due to strong scattering.  (From 
Holley-Bockelmann \etal, in preparation.) }
\label{ss-figure4}
\end{figure}

There are two important issues to be explored here.
 
\begin{itemize}

\item When a spherical (or axisymmetric) model is squeezed adiabatically into
a triaxial configuration, one of the implicit assumptions of adiabatic growth 
is violated.  The evolution of the potential has discretely broken a symmetry 
underlying the second and third integrals of motion, and, incidentally, the 
reversibility of the process is destroyed.  However, the action is still 
conserved, at least approximately, so the \df must bifurcate, leaving an 
excluded region of phase space. In 2-D this region would be forbidden; in 3-D 
other bifurcating branches can cross-over into the newly created vacant region 
of phase space, but do not in general fill it. By construction, this technique 
leaves vacant islands in phase space, and consequently we reach stable and 
stationary triaxial configurations despite the presence of the black hole.  
Numerically these are robust solutions, but they are not guaranteed to be 
robust physically. It is possible that small amounts of relaxation or 
potential fluctuation could rapidly refill these vacated phase-space regions, 
leading to boxlet--chaos transitions, breaking the dynamical equilibrium 
constructed for boxlet--boxlet transitions. Some of this is seen through chaos
induced by numerical scattering in the models.

We cannot yet be sure that our triaxial adiabatic solutions are physically 
robust ones achieved by natural systems, as distinct from mathematical 
curiosities; nevertheless, they are potentially very interesting solutions for 
triaxial systems.

\item We also do not understand well how a transition to chaos occurs in these 
systems.  The scattering by the central singular potential, in and of itself, 
need not induce chaos.  After all, perturbed orbits in Keplerian potentials 
are regular.  Some insight can be gained by considering a toy 2-D model (to be 
compared with the dynamics in the principal plane of a triaxial system).  

Following Devaney (1982), consider a homogenous, {\it anisotropic}\ potential 
of degree $k$, $\Phi (r,\psi) \propto r^{-k}$.  There are three special values 
of $k = \{0,1,2\}$, the first two corresponding to the isothermal and Keplerian 
potential, respectively.  We consider the characteristic exponents in the 
linearised theory for 2-D orbits and explore the effects of varying $k$ and 
the degree of anisotropy.  Solving for the characteristic exponent $\lambda$,
we find
 
\begin{equation}
\lambda (k,r) = {1\over 2} [ ({1\over 2}k - 1)v_r - [ ({1\over 2}k - 1)^2 v_r^2 - 4\Phi''(\psi)]^{1/2} ]
\end{equation}

\noindent 
where $\Phi''$ is the second angular derivative of the potential, which is 
by hypothesis self-similar (i.e., the shape of the isopotential contours is 
independent of radius).  Clearly, realistic galaxy models are not homogenous, 
but at large and small radii they well approximate a homogenous potential, and 
the dynamics, in particular the orbit divergences, are due to {\it local}\ 
potential gradients.

Instability formally occurs when $\lambda$ is imaginary, so the critical 
points occur when $ [({1\over 2}k - 1)^2 v_r^2 - 4\Phi''(\psi)] = 0$.  Using 
Poisson's equation, $\Phi'' = 4\pi r^2 \rho - 2v_c^2 - k(k-1)\Phi$, where, 
by hypothesis, $\Phi''(\psi)$ is independent of radius. We therefore conclude,
that for Keplerian and isothermal potentials orbits in this model go chaotic 
at large $r$, not crossing through the center. For $k=2$, appropriate for the 
outer regions of galaxy models, orbits that are chaotic anywhere are chaotic 
everywhere.

If the dynamics of the toy 2-D homogenous model are a good indicator, then the 
onset of chaos in triaxial models occurs not at small radii, but in the 
transition region between the inner cusp and the region outside $r_h$.
\end{itemize}

\subsection{Uses of triaxial models}

Much work remains to be done.  The following are some implications worth 
exploring.

\begin{itemize}
\item{}The gas dynamics on intermediate scales and possibilities for AGN 
fueling may depend strongly on even mild triaxialities on small scales.

\item{} The dynamical evolution of merging dwarf galaxies and the fate of 
low-mass black holes merging with massive galaxies may be sensitive to 
triaxiality persisting as the central black hole becomes more massive.  
Dynamical friction processes may be expedited through centrophilic orbits in 
triaxial potentials.

\item{} On small scales, triaxiality inside $r_h$ may promote rapid loss-cone 
refilling and be critical for sustaining high tidal-disruption rates and the 
influx of low-mass compact objects coalescing with the central massive black 
hole (Sigurdsson 2003).

\end{itemize}

\section{Conclusions}

Adiabatic growth models provide valuable physical insights into the dynamical
processes and interactions of central massive black holes with their 
surroundings.  Over the last 30 years, a broad range of physically robust 
results on the dynamical influence of the black hole on the surrounding 
stellar population have contributed to our confidence in the reality of 
supermassive black holes and helped provide strong quantitative constraints on 
black hole masses.  Additional physical insight has been gained in 
understanding the ever-fascinating subtleties of Newtonian dynamics of 
many-body systems.  New generations of $N$-body models will allow a broader 
exploration of more realistic, less symmetric systems, more tests of stability 
and secular evolution, and possibly a deeper understanding of dynamical 
processes on small and large scales.

%\smallskip

{\it Acknowledgements:} The author gratefully acknowledges the support of the 
NSF under grant PHY-0203046, the Center for Gravitational Wave Physics at Penn 
State, an NSF-supported Physics Frontier Center, and the hospitality of 
Carnegie Observatories. I would like to thank my collaborators Gerry Quinlan, 
Lars Hernquist, Roeland van der Marel, Chris Mihos, Colin Norman, and 
especially Kelly Holley-Bockelmann for her hard work and for letting me use 
unpublished results.

\end{document}